%% file: main.tex
\pdfoutput=1
\documentclass[11pt]{article}

\usepackage{acl}

\usepackage{times}
\usepackage{latexsym}
\usepackage{changes}

\usepackage[T1]{fontenc}
\usepackage[utf8]{inputenc}

\usepackage{microtype}

\usepackage{inconsolata}

\usepackage{colortbl}
\usepackage{multirow}
\usepackage{graphicx}
\usepackage{subcaption}
\usepackage{stfloats}
\usepackage{tcolorbox}
\usepackage{array}
\usepackage{adjustbox}
\usepackage{booktabs}
\usepackage{graphicx}
\tcbuselibrary{most}
\usepackage{listings} 
\usepackage{xcolor}
\usepackage{xspace}
\usepackage{color}
\usepackage{url}
\usepackage{listings}
\usepackage{ifthen}
\usepackage{enumitem}
\usepackage{makecell}
\usepackage{float}
\usepackage{threeparttable}
\usepackage{diagbox}
\usepackage{amsmath}
\usepackage{cleveref}
\usepackage{lipsum}
\usepackage[ruled,vlined]{algorithm2e}
\usepackage{algorithmic}
\usepackage{colortbl}
\definecolor{bg}{rgb}{0.98, 0.98, 0.98} 
\newcommand{\todoorange}[1]{\textcolor{orange}{~#1}}
\newcommand{\todomagenta}[1]{\textcolor{magenta}{~#1}}
\definecolor{mygreen}{HTML}{238b45}
\newcommand{\todogreen}[1]{\textcolor{red}{~#1}}
\newcommand{\todoviolet}[1]{\textcolor{violet}{~#1}}

\newcommand{\keepnotes}{true}
\ifthenelse{\isundefined{\keepnotes}}{
\newcommand{\civi}[1]{}
\newcommand{\xiaohu}[1]{}
\newcommand{\zifan}[1]{}
\newcommand{\jiahao}[1]{}
}{
\newcommand{\civi}[1]{\todogreen{[Ming: #1]}}
\newcommand{\xiaohu}[1]{\todoorange{[xiaohu: #1]}}
\newcommand{\zifan}[1]{\todomagenta{[zifan: #1]}}
\newcommand{\jiahao}[1]{\todoviolet{[jiahao: #1]}}
}

\newcommand*{\eg}{e.g.,}

\newcommand*{\ie}{i.e.,}

\newcommand{\tabincell}[2]{\begin{tabular}{@{}#1@{}}#2\end{tabular}}

\title{Generalization-Enhanced Code Vulnerability Detection via Multi-Task Instruction Fine-Tuning}

\author{
Xiaohu Du$^{1}\thanks{\ \ National Engineering Research Center for Big Data Technology and System, Services Computing Technology and System Lab, HUST, Wuhan, 430074, China}\thanks{\ \ Hubei Engineering Research Center on Big Data Security, Hubei Key Laboratory of Distributed System Security, HUST, Wuhan,
430074, China}$,
Ming Wen$^{1}$\footnotemark[1]\footnotemark[2]$\thanks{\ \  JinYinHu Laboratory, Wuhan, 430077, China}\thanks{\ \  Corresponding author}\hspace{0.2em}$, 
Jiahao Zhu$^{1}$\footnotemark[1]\footnotemark[2],
Zifan Xie$^{1}$\footnotemark[1]\footnotemark[2],
Bin Ji$^{3}$,
Huijun Liu$^{3}$,\\
{\bf Xuanhua Shi$^{2}$\footnotemark[1]\thanks{\ Cluster and Grid Computing Lab, HUST, Wuhan,
430074, China}, 
Hai Jin$^{2}$\footnotemark[1]\footnotemark[5]}\\
$^1$School of Cyber Science and Engineering, Huazhong University of Science and Technology\\
$^2$School of Computer Science and Technology, Huazhong University of Science and Technology\\
$^3$College of Computer, National University of Defense Technology\\
{\tt \{xhdu, mwenaa, m202271745, xzff, xhshi, hjin\}@hust.edu.cn},\\
{\tt \{jibin, liuhuijun\}@nudt.edu.cn}\\
}
\begin{document}
\maketitle

\input{abstract.tex}
\input{introduction.tex}
\input{related_work.tex}
\input{methodology.tex}
\input{experiments.tex}

\input{results.tex}

\input{conclusion.tex}
\input{limitation}
\input{acknowledgements}

\bibliography{custom}

\appendix

\input{appendix}

\end{document}

%% file: abstract.tex
\begin{abstract}
\textit{Code Pre-trained Models} (CodePTMs) based vulnerability detection have achieved promising results over recent years.
However, these models struggle to generalize as they typically learn superficial mapping from source code to labels instead of understanding the root causes of code vulnerabilities, resulting in poor performance in real-world scenarios beyond the training instances.
To tackle this challenge, we introduce VulLLM, a novel framework that integrates multi-task learning with \textit{Large Language Models} (LLMs) to effectively mine deep-seated vulnerability features.
Specifically, we construct two auxiliary tasks beyond the vulnerability detection task. 
First, we utilize the vulnerability patches to construct a vulnerability localization task.
Second, based on the vulnerability features extracted from patches, we leverage GPT-4 
to construct a vulnerability interpretation task.
VulLLM innovatively augments vulnerability classification by leveraging
generative LLMs to understand complex vulnerability patterns,
thus compelling the model to capture the root causes of vulnerabilities rather than overfitting to spurious features of a single task. 
The experiments conducted on six large datasets demonstrate that VulLLM surpasses seven state-of-the-art models in terms of effectiveness, generalization, and robustness.
\end{abstract}

%% file: introduction.tex
\section{Introduction}
\textit{Code Pre-trained Models} (CodePTMs) such as CodeBERT~\cite{DBLP:conf/emnlp/FengGTDFGS0LJZ20}, GraphCodeBERT~\cite{DBLP:conf/iclr/GuoRLFT0ZDSFTDC21}, and UniXcoder~\cite{DBLP:conf/acl/GuoLDW0022} have been increasingly applied to automated code vulnerability detection over recent years, achieving \textit{state-of-the-art} (SOTA) results. 
In particular, these CodePTMs take code snippets as inputs and predict whether potential vulnerabilities exist in the code.
However, a recent study~\cite{DBLP:conf/emnlp/DuZKZZ23} has highlighted a critical limitation in these models' generalization capabilities, particularly when dealing with \textit{out-of-distribution} (OOD) data.
The limitation arises as existing approaches tend to capture superficial rather than in-depth vulnerability features when learning the mapping from source code to labels.
A notable manifestation is the inability of such approaches to accurately differentiate adversarial examples~\cite{DBLP:conf/emnlp/ZhangMHLXTL23, DBLP:conf/aaai/ZhangLLMLJ20} that merely replace identifiers, indicating that their predictions are affected by factors that are irrelevant to the vulnerability. 
Furthermore, the learning paradigm via mapping from source code to labels struggles with the generalization ability when handling vulnerable code from multiple projects, as the code from different projects often varies in programming style and application contexts, thus leading to divergent distributions of vulnerability features~\cite{DBLP:conf/emnlp/DuZKZZ23}.

It is noteworthy that recently emerged \textit{Large Language Models} (LLMs) have demonstrated remarkable reasoning and generalization capabilities~\cite{DBLP:journals/corr/abs-2304-04370, DBLP:journals/corr/abs-2311-13982} across various domains, thus inspiring us to harness them for developing more robust vulnerability detection models. 
However, directly applying LLMs in code vulnerability detection encounters various challenges due to the absence of specialized training tailored to this particular task~\cite{DBLP:journals/corr/abs-2308-12697, DBLP:journals/corr/abs-2311-12420}. 
%\civi{Refine the above challenge.}\xiaohu{update}
Unfortunately, simply employing methods similar to CodePTMs to fine-tune LLMs will still introduce the above issues with generalization.
To tackle these challenges with LLMs, this study presents VulLLM, a novel technique that integrates code vulnerability knowledge into LLMs through instruction tuning~\cite{DBLP:journals/corr/abs-2308-10792}.

To prevent the model from learning spurious features, we employ the multi-task learning paradigm to enable LLMs to learn deep-seated features rather than spurious ones. 
The insight is to enhance the mapping from source code to labels by adding two auxiliary tasks which aim to gain a deeper understanding of vulnerabilities by identifying their root causes and locating the corresponding vulnerable code elements.
The first auxiliary task, vulnerability localization, identifies vulnerable code elements (\eg~statement) that are extracted from the patch.
The second task, vulnerability interpretation, identifies vulnerabilities' root causes and outputs their textual interpretation.
As no ready-made interpretation exists, we generate them using GPT-4.
However, LLMs still face challenges in vulnerability detection, let alone identifying the root causes of vulnerabilities. 
In a manual assessment, ChatGPT barely understands half of the vulnerabilities it `detects' based on simple prompts~\cite{DBLP:journals/corr/abs-2308-12697}.
To tackle this challenge, we introduce the patch-enhanced \textit{Chain-of-Thought~\cite{DBLP:conf/nips/Wei0SBIXCLZ22} with Self-Verification} (CoT-SV), which demonstrates effective performance to avoid error accumulation and illusions in CoT, thus enhancing the reliability of LLMs~\cite{DBLP:conf/icml/Ni0RSYWL23, DBLP:journals/corr/abs-2305-11738}.
The validations in CoT-SV include vulnerability labels, \textit{Common Vulnerabilities and Exposures} (CVE) descriptions, and vulnerability lines with contexts extracted from the patch based on \textit{Program Dependency Graph} (PDG)~\cite{DBLP:journals/tdsc/LiSysevr}.
%\civi{add a recent work that uses PDG. 1987 is very old.} \jiahao{update}
Auxiliary tasks enhance the variety and depth of features (\ie~ the vulnerable location and root cause), thereby improving the model's comprehension of the domain knowledge of code vulnerabilities. 
Moreover, the diversity of features contributes variably across various tasks, compelling the model to seek solutions that perform well across all tasks, thereby preventing overfitting to specific spurious features of a single task.

In addition to the above data generated for multi-task learning, our training data also includes two real-world vulnerability datasets with the highest label accuracy from two manual evaluations~\cite{DBLP:conf/raid/0001DACW23, DBLP:conf/icse/CroftBK23}: Devign~\cite{DBLP:conf/nips/ZhouLSD019} and DiverseVul~\cite{DBLP:conf/raid/0001DACW23}, to further enhance LLMs' learning of various code vulnerabilities.  
Furthermore, to verify the extensive applicability of our framework, we select three widely-used foundational models to construct VulLLM, including a general LLM, Llama-2~\cite{DBLP:journals/corr/abs-2307-09288}, alongside two CodeLLMs, namely CodeLlama~\cite{DBLP:journals/corr/abs-2308-12950} and StarCoder~\cite{DBLP:journals/corr/abs-2305-06161}.
The results indicate that VulLLM outperforms seven existing SOTA vulnerability detection models.
Overall, compared to the best baseline, UniXcoder, VulLLM demonstrates superior effectiveness with an improvement in F1 score by 8\% across six datasets. 
Notably, within these datasets, the F1 score of VulLLM has increased by 8.58\% on four OOD datasets, indicating its better generalization. 
Furthermore, we introduce three adversarial attacks to verify the robustness of these models.
Under these attacks, the overall F1 score of VulLLM improves by 68.08\% compared to UniXcoder, highlighting its enhanced robustness.

Our contributions are summarized as follows:
\vspace{-1.5mm}
\begin{itemize}[leftmargin=*]
    \vspace{-1 mm}
    \item \textbf{Idea.} We propose a novel perspective to leverage the interpretability of GPT-4 for generating vulnerability interpretation to enhance the vulnerability understanding of LLMs.
    \vspace{-1mm}
    \item \textbf{Approach.} We propose VulLLM, a framework to detect vulnerabilities with LLMs through multi-task instruction-tuning. To our best knowledge, it is the first attempt to use instruction-tuned LLMs for vulnerability detection.
    \vspace{-1mm}
    \item \textbf{Evaluation.} We conduct extensive experiments across six datasets and find that our approach significantly improves the effectiveness, generalization and robustness in detecting vulnerabilities. We also release the code and data at: \url{https://github.com/CGCL-codes/VulLLM}.
    
\end{itemize}

%% file: related_work.tex
\section{Related Work}
\subsection{Code Vulnerability Detection}
Code vulnerability detection is of significant importance for the secure and stable operation of software systems.
Deep learning methods can automatically learn and generalize features of vulnerabilities from extensive code samples, enabling automated inference of vulnerability patterns. 
This paradigm has gained widespread attention over recent years. 
Early vulnerability detection methods utilize \textit{Graph Neural Networks} (GNNs) to learn vulnerability features. 
With the development of the Transformer~\cite{DBLP:conf/nips/VaswaniSPUJGKP17} architecture, many CodePTMs, such as CodeBERT~\cite{DBLP:conf/emnlp/FengGTDFGS0LJZ20}, GraphCodeBERT~\cite{DBLP:conf/iclr/GuoRLFT0ZDSFTDC21}, and UniXcoder~\cite{DBLP:conf/acl/GuoLDW0022}, have achieved better performance in vulnerability detection. 
They are mainly pre-trained on extensive datasets with code and text, and have demonstrated outstanding performance in multiple code-related downstream tasks.
Additionally, some approaches are built upon CodePTMs. 
ReGVD~\cite{DBLP:conf/icse/NguyenNNLTP22} encodes source code as a graph with nodes representing code tokens and features initialized based on CodePTMs.
EPVD~\cite{DBLP:journals/tse/ZhangLHXL23} divides the code into various execution paths based on \textit{Control Flow Graph} (CFG) and learns different path representations based on CodePTMs. 
These approaches mainly train models via learning from a single-task. 
In this work, we utilize the paradigm of multi-task learning to enable LLMs to better understand code vulnerabilities.

\input{figures/overview.tex}
\subsection{Self-Verification in LLM}
CoT~\cite{DBLP:conf/nips/Wei0SBIXCLZ22} is a prompting technique  to solve problems with LLMs, which employs a series of reasoning steps to tackle complex issues, akin to the thought process of humans in solving problems. 
Self-verification~\cite{DBLP:journals/corr/abs-2308-03188} aims to mitigate hallucinations~\cite{DBLP:conf/acl/LinHE22} and unfaithful reasoning in LLMs~\cite{DBLP:conf/iclr/GolovnevaCPCZFC23, DBLP:journals/corr/abs-2301-13379}, while reducing error accumulation in CoT as well~\cite{weng2023large}.
Specifically, it corrects the adverse behaviors of LLMs through feedback, which also aligns with human learning strategies, that is, a cycle of attempting, making mistakes, and correcting.
Verification often comes from two sources: manual validation and automatic validation.
Manual validation tends to be more congruent with human preferences. 
For instance, InstructGPT~\cite{DBLP:conf/nips/Ouyang0JAWMZASR22} improves GPT-3~\cite{DBLP:conf/nips/BrownMRSKDNSSAA20} through human feedback. 
Automatic validation can originate from the LLM itself or external knowledge.
For instance, SelfCheck~\cite{DBLP:journals/corr/abs-2308-00436} demonstrates LLM's ability for correcting errors in CoT independently, without external resources. 
The results from different stages are employed to derive an overall confidence score, which is subsequently utilized as a weight to cast votes among multiple solutions to the same problem, thereby enhancing the accuracy of the response.
Validations based on external knowledge can originate from various sources, such as Wikipedia~\cite{DBLP:journals/corr/abs-2307-03987} and search engines~\cite{DBLP:journals/corr/abs-2305-11738}. 
In this study, we obtain validation for vulnerability interpretation from the  features extracted from the corresponding vulnerability patches (see Section~\ref{sec:feature} for more details).

%% file: figures/overview.tex
\begin{figure*}[t!]
	\centering
        \setlength{\belowcaptionskip}{-10pt}
	\includegraphics[width=1.0\linewidth]{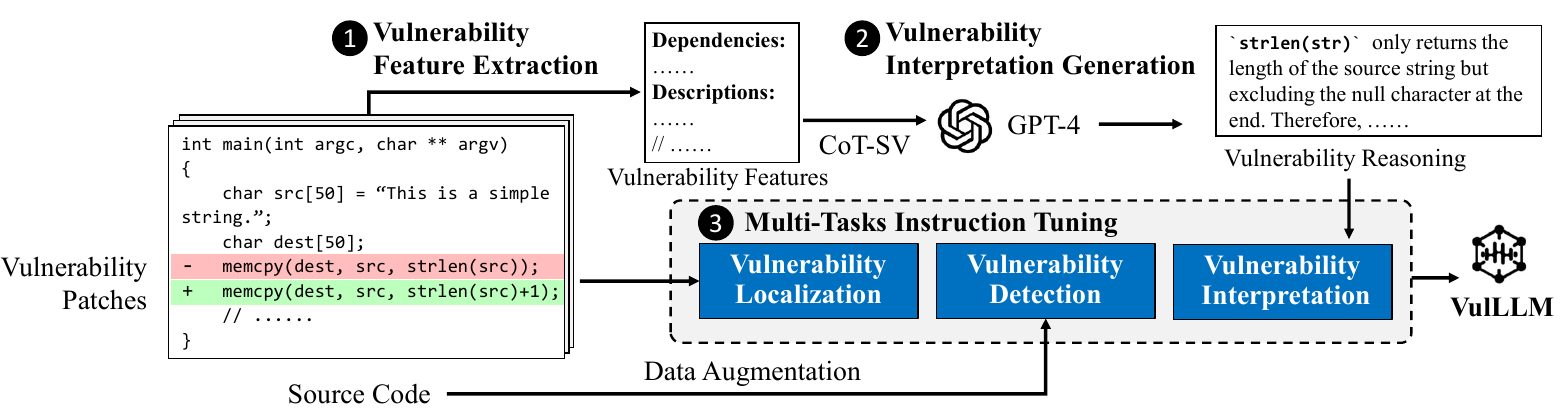}
	\centering
	\vspace{-5mm}
	\caption{The general workflow of VulLLM}
	\label{fig:workflow}
\end{figure*} 

%% file: methodology.tex
\section{Methodology}
\subsection{Overview}
Figure~\ref{fig:workflow} illustrates an overview of VulLLM, which comprises three main components: vulnerability features extraction (Section~\ref{sec:feature}), vulnerability interpretation generation (Section~\ref{sec:interpretation}), and multi-task instruction fine-tuning (Section~\ref{sec:tuning}). 

\subsection{Vulnerability Features Extraction}\label{sec:feature}
\input{figures/feature_new}
Vulnerability features serve as the essential cues for vulnerability interpretation. 
In this study, we aim to enhance the generalization of LLM in the context of vulnerabilities. 
Specifically, we explore the use of \textit{vulnerability lines}, \textit{vulnerability context}, and \textit{CVE descriptions} as potential cues for understanding vulnerabilities.

\noindent\textbf{Vulnerability Lines}. Vulnerability lines directly point out the vulnerable code elements. 
We extract vulnerability lines from the patches of the vulnerable code. 
Patches are generated to fix existing vulnerabilities via adding or deleting certain code elements. 
Following existing study~\cite{DBLP:journals/ese/NguyenDM16}, we consider the deleted lines in patches to directly reflect the vulnerable semantics.
For instance, if the deleted lines involve unsafe coding practices, such as \textit{improper memory management} or \textit{insecure input validation}, these lines could be the direct root cause of the vulnerability. 

\noindent\textbf{Vulnerability Context}. Vulnerability context refers to the surrounding code (\eg~conditions, checkers, \textit{etc.}) that provide a broader understanding of a security vulnerability. 
Typically, we extract code statements that have direct or indirect data dependencies and control dependencies along with the vulnerable lines as vulnerability context.
To extract these code statements, we first use JOERN~\cite{JOERN} to generate the Program Dependency Graph (\ie~PDG)~\cite{DBLP:journals/tdsc/LiSysevr} for the vulnerability functions.
PDG is a directed acyclic graph where nodes represent code elements, and various types of directed edges between nodes represent relationships between code elements (e.g., if there exists a data dependency edge originating from node A and directed towards node B, it indicates that node B depends on a data variable defined at node A).
PDG has been widely utilized in the domain of vulnerability detection~\cite{DBLP:conf/sigsoft/Li0N21,DBLP:journals/tse/ZhangLHXL23}.
Specifically, we start from the nodes corresponding to the vulnerabilities and identify neighboring nodes within a $k$-hop distance through both data dependency edges and control dependency edges (either outgoing or incoming). The code lines corresponding to these nodes are then added to the vulnerability context. 
Figure~\ref{fig:feature} illustrates an example of vulnerability CVE-2018-7751, which belongs to the type of CWE-835. The left side depicts the applied patch, the middle section showcases the corresponding PDG, and the right side displays the extracted vulnerability features.
For the vulnerability line at line 9, there is a control dependency edge from line 8 to line 9, as well as three data dependency edges—one from line 3 to line 9, one from line 9 to line 14 and another from line 9 to line 16. If k is set to 1, the contextual scope encompasses statements found at lines 8, 3, 14, and 16.
The parameter $k$ is used to control the length of the generated vulnerability context, as dependency relationships in real-world code can be particularly complex. In our implementation, the value of $k$ is set to 1, considering the limited input length capacity of LLMs.

\noindent\textbf{CVE Descriptions.} This information shed lights on the root causes of vulnerabilities, as they comprehensively detail common security weaknesses in software and hardware.
These descriptions are invaluable for understanding the root causes of vulnerabilities, which often offer relevant background and context, explaining how these weaknesses come about and how they might be exploited under different circumstances.
To collect the CVE descriptions, we have scraped them for each CVE from the NVD~\cite{NVD}.

\subsection{Vulnerability Interpretation Generation}\label{sec:interpretation}
\input{figures/CoT_SV.tex}
The vulnerability features extracted in the previous section serve as the critical information for validating the output of each step in CoT. 
Figure~\ref{fig:CoT} illustrates the implementation of CoT-SV.

\noindent\textbf{Step 1.} Given the demonstrated efficacy of role-playing in prompt engineering~\cite{DBLP:journals/corr/abs-2308-07702,DBLP:journals/corr/abs-2305-16367}, our initial step involves adopting a role-centric prompting strategy, specifically focusing on vulnerability detection, to ensure that the model remains concentrated on the task throughout the workflow.
We use the following prompt as adapted from a prior work~\cite{DBLP:journals/corr/abs-2308-12697} in this step. 
\vspace{-2mm}
\begin{figure}[H]
	\centering
    \setlength{\belowcaptionskip}{-10pt}
	\includegraphics[width=0.95\linewidth]{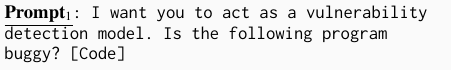}
	\centering
	\vspace{-2mm}
\end{figure} 

\noindent where [Code] refers to the potentially buggy code.
Previous study indicates that the accuracy of LLM under such prompt is suboptimal~\cite{DBLP:journals/corr/abs-2304-07232}. 
Fortunately, we have the ground truth for each code.
Therefore, if the LLM produces an incorrect output, we can address it in subsequent steps.
More importantly, this initial step serves to reinforce LLM's acknowledgment of the presence of vulnerabilities, facilitating subsequent vulnerability reasoning.
To prevent overfitting, we execute \textbf{Step 1} for an equal number of vulnerable and non-vulnerable code examples to generate vulnerability explanations. 
For non-vulnerable code, GPT-4 will provide interpretations indicating the absence of vulnerabilities, which will be used to construct the dataset for non-vulnerable code in the vulnerability explanation task.
For vulnerable code, more precise vulnerability interpretations will be obtained through continuous verification of GPT-4's outputs in subsequent steps.

\noindent\textbf{Step 2 - Step 4.} We adapt a unified prompt template for various vulnerability features based on existing Self-Verification templates~\cite{DBLP:journals/corr/abs-2308-03188, ling2023deductive}. 
The validity of the output from each individual step is verified using a directive composed of the following components: (1) information required to be verified for the current step. (2) an instruction for validity verification, such as \textit{Please double-check the answer and analyze its correctness.}~\cite{ling2023deductive} (3) requirements for the output of the subsequent step under the LLM. Based on the above design, the prompts for obtaining vulnerability interpretations are as follows:
\vspace{-2mm}
\begin{figure}[H]
	\centering
    \setlength{\belowcaptionskip}{-10pt}
	\includegraphics[width=0.95\linewidth]{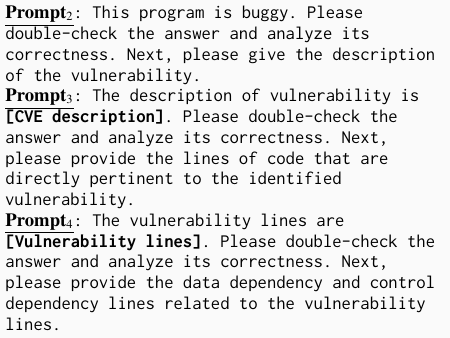}
	\centering
	\vspace{-2mm}
\end{figure}

\noindent where [CVE description] and [Vulnerability lines] denote the vulnerability features as extracted in the previous subsection.

\noindent\textbf{Step 5.} Vulnerability context constitutes the final features for verification within CoT-SV.
Following this verification, we employ the LLM to synthesize the vulnerability interpretation by integrating the aforementioned categories of features.
Specifically, we instruct the LLM to generate a vulnerability interpretation that refers to the vulnerability lines and vulnerability context.
\vspace{-1mm}
\begin{figure}[H]
	\centering
    \setlength{\belowcaptionskip}{-10pt}
	\includegraphics[width=0.95\linewidth]{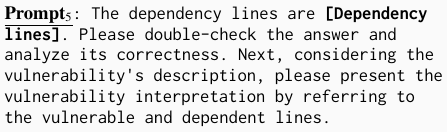}
	\centering
	\vspace{-2mm}
\end{figure} 

\vspace{-1mm}
The results generated by CoT-SV encompass rich knowledge specific to vulnerabilities. 
For instance, the CVE description constrains the high-level overview of the vulnerability, whereas the vulnerability lines pinpoint its precise location. 
Furthermore, the dependency lines characterize the context of vulnerabilities.
Such knowledge ensures that the reasoning process integrates an extensive repository of domain-specific knowledge on vulnerabilities.
Concurrently, CoT-SV further abstracts the data flow and control flow of vulnerabilities, translating the dependency lines into natural language descriptions of the vulnerability context. 
Finally, all the generated data are manually verified to exclude instances where the final judgment of GPT-4 is still incorrect.
We take the vulnerable code in Figure~\ref{fig:feature} as an example and present different interpretations of \textbf{Step 1} and the entire CoT-SV generation process separately in Appendix~\ref{sec:cot-sv}.

\subsection{Multi-Task Instruction Fine-tuning}\label{sec:tuning}
\textbf{Data Preparation.}
The data serving for the above multi-task data originates from the PatchDB~\cite{DBLP:conf/dsn/WangWF0J21},
which contains the patch information, a feature not commonly found in other datasets. 
The diffs in patches are directly associated with vulnerabilities and essential for obtaining vulnerability interpretations. 
However, this patch-based vulnerability dataset is limited in its quantity and insufficient for LLMs to learn a broad range of vulnerability patterns. 
To infuse LLMs with more vulnerability knowledge, we further incorporate two additional datasets for the vulnerability detection task: DiverseVul~\cite{DBLP:conf/raid/0001DACW23} and Devign~\cite{DBLP:conf/nips/ZhouLSD019}, which are the two real-world datasets with the highest label accuracy in manual evaluation sampling~\cite{DBLP:conf/icse/CroftBK23, DBLP:conf/raid/0001DACW23}. 
Increasing the data volume for the primary task ensures that the model does not deviate from the core objective while learning auxiliary tasks, which also helps in maintaining the priority of the primary task. 
These two datasets contain 797 distinct projects. 
A common challenge when training on multi-source code datasets is how to handle the variation of feature distribution among different projects.
To mitigate this variation, we employ random identifier substitution to enhance the generalization of LLMs on multi-project code. 
The core idea is to reduce model dependence on a specific project's features by increasing data diversity, thereby minimizing the risk of overfitting and enhancing the model's adaptability to different coding styles. 
Specifically, we replace 10\% of the existing identifiers within the original code with randomly chosen identifiers sourced from the complete dataset.
The input for all three tasks is the source code. 
For the output, vulnerability detection yields a label of 0 or 1.
Vulnerability localization identifies the vulnerable line as extracted in Figure~\ref{fig:feature}, and vulnerability interpretation provides the natural language (the final result of CoT-SV, as demonstrated in Appendix~\ref{sec:cot-sv}).

\noindent\textbf{Instruction Fine-tuning.} Instruction fine-tuning aims to optimize the response of LLMs to specific directives, thus ensuring the alignment with the requirements of a particular task.
Specifically, we employ instruction fine-tuning to train a more specialized, adaptable, and efficient LLM for vulnerability detection.
For each task, we provide a distinct instruction. 
By integrating this instruction with the input code, the LLM is capable of producing specific outputs. 
Subsequently, the LLM quantifies the discrepancy between the generated output and the anticipated target, leveraging this deviation to fine-tune the weights of LLM.
In this work, we adapt the template provided by Alpaca~\cite{taori2023alpaca} for instruction fine-tuning: 
\vspace{-1mm}
\begin{figure}[H]
	\centering
    \setlength{\belowcaptionskip}{-10pt}
	\includegraphics[width=0.95\linewidth]{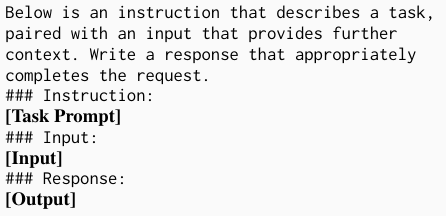}
	\centering
	\vspace{-2mm}
\end{figure}

\noindent where [Input] and [Output] are obtained from the above \textbf{Data Preparation}, and the [Task Prompt] directs LLMs to generate task-specific outputs based on different tasks. 
We provide specific examples of instruction data for different tasks in Appendix~\ref{sec:cot-sv}.

%% file: figures/feature_new.tex
\begin{figure*}[t!]
	\centering
        \setlength{\belowcaptionskip}{-10pt}
	\includegraphics[width=0.95\linewidth]{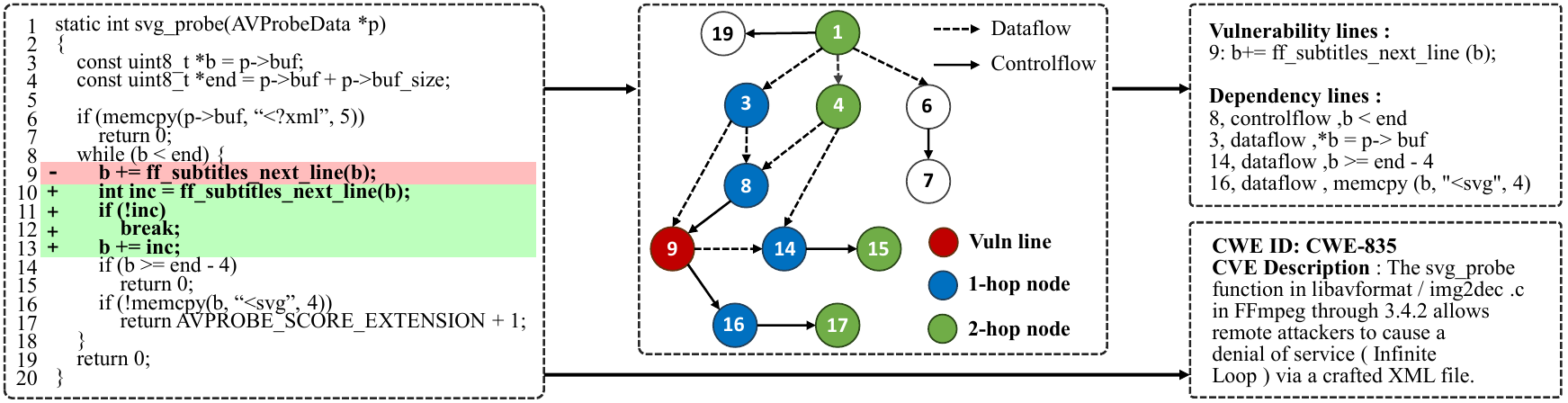}
	\centering
	\vspace{-3mm}
	\caption{An example of vulnerability feature extraction}
	\label{fig:feature}
\end{figure*} 

%% file: figures/CoT_SV.tex
\begin{figure}[t!]
	\centering
    \setlength{\belowcaptionskip}{-10pt}
	\includegraphics[width=0.95\linewidth]{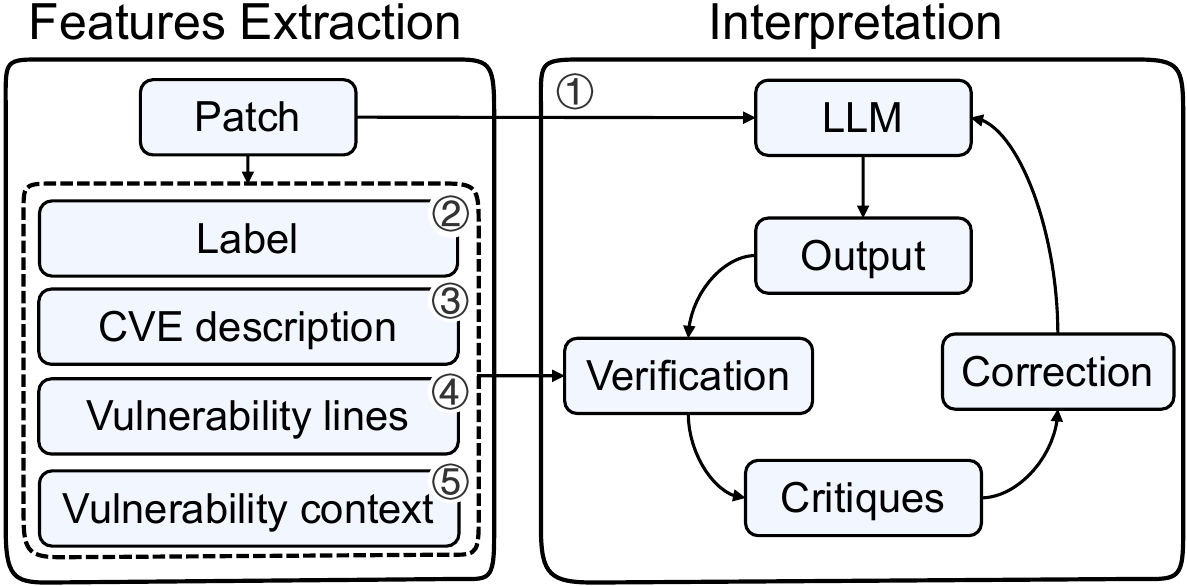}
	\centering
	\vspace{-2mm}
	\caption{The implementation of CoT with Self-Verification.  Numbered circles denotes the five steps}
	\label{fig:CoT}
\end{figure} 

%% file: experiments.tex
\section{Experimental Setup}
\subsection{Datasets}
We select six widely-used C/C++ vulnerability detection datasets to evaluate different models extensitively: DiverseVul~\cite{DBLP:conf/raid/0001DACW23}, Devign~\cite{DBLP:conf/nips/ZhouLSD019}, BigVul~\cite{DBLP:conf/msr/FanL0N20}, CVEfixes~\cite{DBLP:conf/promise/BhandariNM21}, ReVeal~\cite{DBLP:journals/tse/ChakrabortyKDR22}, and Juliet~\cite{DBLP:journals/computer/BolandB12}. 
The first two datasets are involved in model training (denoted as \textbf{\textit{$Dataset_1$}}) while the latter four datasets are not presented in the training process (denoted as $Dataset_2$). 
Therefore, the results on $Dataset_2$ can further reflect the models' generalizability besides effectiveness. 
The details of the datasets can be found in Appendix~\ref{sec:dataset}.

\subsection{Baselines}
In our evaluation, we compare VulLLM with the following SOTA models, encompassing diverse architectures and approaches to ensure a broad spectrum of comparison. 
Specifically, our baselines include two GNNs-based models: Devign~\cite{DBLP:conf/nips/ZhouLSD019} and ReVeal~\cite{DBLP:journals/tse/ChakrabortyKDR22}, three CodePTMs: CodeBERT~\cite{DBLP:conf/emnlp/FengGTDFGS0LJZ20}, GraphCodeBERT~\cite{DBLP:conf/iclr/GuoRLFT0ZDSFTDC21}, UniXcoder~\cite{DBLP:conf/acl/GuoLDW0022}, and two models based on CodePTMs: ReGVD~\cite{DBLP:conf/icse/NguyenNNLTP22} and EPVD~\cite{DBLP:journals/tse/ZhangLHXL23} that are specifically designed for vulnerability detection. 
The details of these baselines can be found in Appendix~\ref{sec:baselines}.
The implementation details of all the baselines and our approach can be found in Appendix~\ref{sec:mplementation}.

%% file: results.tex
\section{Results}
\subsection{Effectiveness and Generalization} 
\input{tables/performance}
We select two versions with different parameter sizes of Llama-2, CodeLlama, and StarCoder as our base models respectively, to validate the extensive applicability of our framework. 
Specifically, we evaluate the effectiveness of various approaches across the six test datasets. 
Notably, $Dataset_2$, which is not involved in model training, can further reflect the model's generalization. 

The results shown in Table~\ref{tab:results_on_7_datasets} indicate that VulLLM, based on CodeLlama-13B, exhibits the highest performance, yielding an overall F1 score of 66.54\%.
Across all models and datasets, VulLLM based on CodeLlama-13B consistently ranks either first or second, which outperforms the 7 selected baselines.  
Therefore, our subsequent analysis of VulLLM is based on CodeLlama-13B. 
Compared to the best baseline model, UniXcoder, VulLLM demonstrates an overall effectiveness improvement by 8\% (\ie~(66.54-61.61)/61.61) across all six datasets.
Notably, it exhibits improvements by 8.58\% (\ie~(64.16-59.09)/59.09) in the F1 score on $Dataset_2$, indicating its superior generalization.
In addition, existing models generally exhibit poor generalization ability. 
In the 20 (5 models $\times$ 4 datasets) generalization experiments on CodePTMs, the average F1 score of the baseline models decrease by 11.36\% to 38.36\% (24.33\% on average) compared to $Dataset_1$. 
In contrast, VulLLM demonstrates a much smaller decrease in performance, decreasing by only 10.03\%, and crucially, maintains F1 scores above 60\% across all datasets.
While GNN-based models seem to exhibit a lesser performance decline on $Dataset_2$ compared to CodePTMs, seemingly demonstrating better generalization, their poorer effectiveness and complex data preprocessing make them significantly less versatile than other models, limiting their practical applicability.

\subsection{Robustness}\label{sec:robustness}
\input{tables/adversarial_table}
We select DiverseVul and ReVeal, from $Dataset_1$ and $Dataset_2$ respectively, in this experiment. 
DiverseVul is chosen due to its inclusion of a wide variety of projects, thus offering a comprehensive evaluation of the model's robustness. 
ReVeal is selected because its data is sourced from the Chromium and Debian packages, which differs from other datasets that originate from NVD~\cite{NVD} or GitHub repositories. 
We select two adversarial attacks for CodePTMs that are based on random identifier replacement: MHM~\cite{DBLP:conf/aaai/ZhangLLMLJ20} and WIR-Random~\cite{DBLP:conf/issta/ZengTZLZZ22}, since they achieve the highest attack success rate in recent evaluations~\cite{DBLP:conf/sigsoft/Du0WW023}.
In addition, we also construct an attack based on random dead code insertion, with the form of the dead code derived from DIP~\cite{DBLP:conf/acl/NaC023}.
Since the code of DIP is not publicly available, and our objective is to compare the robustness of models under different attacks rather than pursuing the highest attack success rate, we do not directly use DIP.
The details of these attacks can be found in Appendix~\ref{sec:attack}.

We select UniXcoder, which performs the best over all the baselines in Table~\ref{tab:results_on_7_datasets}, and VulLLM for robustness evaluation.
Table~\ref{tab:adversarial} reveals the F1 score of two models under three adversarial attacks. 
It shows that UniXcoder exhibit a significant decline in performance under various adversarial attacks. 
Notably, the performance under ReVeal is lower than that under DiverseVul, indicating that UniXcoder has poorer robustness on OOD data. 
This reduced robustness is a testament to their insufficient generalization. Conversely, VulLLM demonstrates superior robustness across all attacks. 
Compared to UniXcoder, VulLLM demonstrates an average improvement by 68.08\%. 
More importantly, VulLLM does not show a further decline in robustness on OOD samples, indicating that its robustness and generalization are superior to UniXcoder.
\input{figures/probability_density}

To explore the reasons behind the differences in robustness between the two models, we further examine the probability densities of correct predictions made by VulLLM and UniXcoder on the two datasets.
We particularly emphasize the importance of high prediction probabilities, as accurate probability predictions are especially crucial when performing safety-critical tasks.
As illustrated in Figure~\ref{fig:density}, the overall performance analysis indicates that VulLLM exhibits superior probability density over UniXcoder across two datasets.
On the DiverseVul dataset, VulLLM shows a significantly higher density in the high probability regions (close to 1.0), indicating stronger confidence in its results. 
Its curve peaks around a probability value of approximately 0.9, and then rapidly declines, suggesting a higher concentration in high probability predictions.
On the ReVeal dataset, the difference in density curves between the two models is not as pronounced as in DiverseVul, but VulLLM still maintains a higher density in regions where the probability is greater than 0.8. 
Particularly in the probability range of 0.8 to 1.0, its density curve is above UniXcoder, peaking near a probability of 0.9.
In summary, VulLLM exhibits greater confidence in its predictions, especially when providing high probability forecasts, contributing to its higher robustness.

\input{tables/ablation}
\subsection{Ablation Study}
In this section, we investigate the impact of multi-task learning and data augmentation. 
As demonstrated in Table~\ref{tab:ablation}, after removing multi-task learning  (``w/o MT''), the model exhibits a performance decline across all datasets, with an overall relative reduction by 9.30\%.
This observation underscores the pivotal role of multi-task learning within our approach, evidencing its substantial contribution towards enhancing the model's effectiveness and generalization.
When the data augmentation component is removed (``w/o DA''), the model exhibits a decrease in average performance by 2.98\%.
Across different datasets, the model shows a decline in performance on four datasets, but an increase on ReVeal and Juliet. 
Such variations suggest that while the data augmentation incorporated into VulLLM effectively enhances model effectiveness, it may simultaneously impair the model's generalization on certain datasets. 
Finally, when two components are removed (w/o ``DA\&MT''), its performance decreases across five datasets.
Notably, on the Devign dataset, its performance is even inferior to that of the three CodePTMs. Additionally, on both BigVul and CVEfixes, it falls short of the performance achieved by ReGVD. 
In summary, the ablation study clearly demonstrates the indispensable roles that multi-task learning and data augmentation play in enhancing the model's overall performance. 
Notably, multi-task learning emerges as the more impactful of the two components, playing a pivotal role in enhancing the model's performance.

\subsection{Sensitivity to hyper-parameter}\label{parameter}
\input{tables/max_length}
\noindent \textbf{Auxiliary task samples.} Considering resource constraints and training efficiency, previous models are trained within a context length of 512. 
Consequently, the amount of auxiliary task data included is limited. 
To further explore the impact of the samples of auxiliary tasks on the performance of VulLLM, we expand the training context lengths to 1,024 and 2,048 to include more auxiliary task samples. 
We present results for two representative datasets, similar to Section~\ref{sec:robustness}, and list the average values for six datasets, as shown in Table~\ref{tab:example}.
We find that an increased number of auxiliary task samples generally leads to a slight improvement in model performance, especially on OOD samples. 
However, there is a noticeable decline in the model's performance on in-distribution samples. 
The changes can be attributed to multi-task learning, where a model learns various tasks together, focusing on features common to all tasks. With more auxiliary task samples, the model adapts to diverse data, improving its overall applicability. However, this broad focus might lead to less optimal performance on specific tasks, as the model might miss finer, unique features of the original training set.
\input{figures/ranks}

\noindent \textbf{Training parameters.}
To demonstrate the sensitivity of VulLLM to training parameters, we conduct experiments with different ranks in LoRA, which are proportional to the training parameters.
As shown in Figure~\ref{fig:ranks}, we observe that the average F1 score increases as the rank increases, reaching its peak at 16. 
Further increasing the rank value leads to a decrease in performance, following a trend similar to existing works~\cite{DBLP:conf/iclr/HuSWALWWC22}.
This phenomenon may be attributed to a limited number of training parameters constraining the model's learning capacity, while an excessive number of parameters may lead to overfitting or excessive complexity in handling the parameters.

%% file: tables/performance.tex
\begin{table*}[htb]
\centering
\small
\vspace{-2mm}
\renewcommand{\arraystretch}{1.1}
\setlength\tabcolsep{4.0pt}
\begin{tabular}{lr|cc|cccc|ccc}
\toprule
\multirow{2}{*}{Methods} &\multicolumn{1}{c|}{\multirow{2}{*}{Size}} &\multicolumn{2}{c|}{$Dataset_1$} &\multicolumn{4}{c|}{$Dataset_2$} &\multicolumn{3}{c}{Average}   \\ 
&&DiverseVul &Devign &BigVul &CVEfixes &ReVeal &Juliet &$Dataset_1$ &$Dataset_2$ &All \\
\midrule  
Devign &1M &60.36 &58.93 &53.14 &55.09 &59.43 &57.71 &59.65 &56.34 &57.44 \\
ReVeal &1M &57.04 &53.84 &53.92 &53.49 &56.67 &57.86 &55.44 &55.49 &55.47  \\
\midrule  
CodeBERT &125M &65.45 &66.81 &55.19 &56.03 &46.48 &\multicolumn{1}{r|}{5.36}  &66.13 &40.77 &49.22     \\
GraphCodeBERT &125M &66.00 &66.62 &54.70 &58.82 &55.61 &31.22 &66.31 &50.09 &55.50     \\      
UniXcoder &126M &67.27 &66.05 &56.06 &59.35 &59.57 &61.36 &66.66 &59.09 &61.61     \\ 
\midrule  
ReGVD  &125M &65.86 &61.14 &58.55 &60.86 &54.94 &36.87 &63.50 &52.81 &56.37      \\
EPVD          &125M &66.85 &66.76 &52.89 &57.73 &48.95 &25.86 &66.81 &46.36 &53.17      \\
\midrule  
\multirow{2}{*}{VulLLM-L2}  
&7B &65.31   &66.24 &57.84 &55.47   &52.80 &64.08 &65.78 &57.55  &60.29    \\  
&13B &\underline{70.42}   &\underline{70.29} &59.54 &61.48   &57.79 &57.04 &\underline{70.36} &58.96 &62.76    \\
\multirow{2}{*}{VulLLM-SC}  
&7B &60.43   &63.46 &62.31 &\textbf{64.50}   &50.81 &63.32 &61.95 &60.24 &60.81    \\  
&15B &62.02   &62.77 &46.17 &52.09   &59.48 &\textbf{69.50} &62.40 &56.81 &58.67    \\

\multirow{2}{*}{VulLLM-CL}  &7B &68.23   &67.84 &\underline{63.51} &59.26   &\textbf{66.45} &58.84 &68.04 &\underline{62.02} &\underline{64.02}    \\  
&13B & \textbf{70.99}   & \textbf{71.63} &\textbf{64.42} & \underline{61.92}   &\underline{64.52} &\underline{65.77} &\textbf{71.31} &\textbf{64.16} &\textbf{66.54}    \\ 
\bottomrule
\end{tabular}
\caption{\label{tab:results_on_7_datasets}The F1 scores on six datasets. The abbreviations ``L2'', ``SC'', and ``CL'' refer to the Llama-2, StarCoder, and CodeLlama, respectively. 
The best results are highlighted in bold, while the next best results are  underlined}
\vspace{-3mm}
\end{table*}

%% file: tables/adversarial_table.tex
\begin{table}[htb]
  \centering
  \small
  \renewcommand{\arraystretch}{1.2}
  \setlength\tabcolsep{1.2mm}
  
\begin{tabular}{cc|cc|c}
\toprule
Attack & Model & DiverseVul & ReVeal &Total Avg \\
\midrule 
\multirow{2}{*}{MHM}    & UniXcoder &24.97	&20.48	&22.73 \\ 
&VulLLM    &\textbf{33.22}	&\textbf{40.06}	&\textbf{36.64} \\ 
\midrule 
\multirow{2}{*}{WIR}    & UniXcoder &\ \ 4.42	&\ \ 2.18	&\ \ 3.30 \\ 
&VulLLM    &\textbf{16.91}	&\textbf{25.25}	&\textbf{21.08} \\ 
\midrule 
\multirow{2}{*}{DCI}    & UniXcoder &37.78	&31.94	&34.86 \\ 
&VulLLM    &\textbf{42.32}	&\textbf{46.94}	&\textbf{44.63} \\ 
\bottomrule

\end{tabular}
\caption{\label{tab:adversarial}The F1 scores under adversarial attacks. For simplicity, DCI denotes the Dead Code Insertion attack}
\vspace{-3mm}
\end{table}

%% file: figures/probability_density.tex
\begin{figure}[htbp]
\centering
\begin{subfigure}{0.23\textwidth}  
  \centering
    \includegraphics[width=1\linewidth]{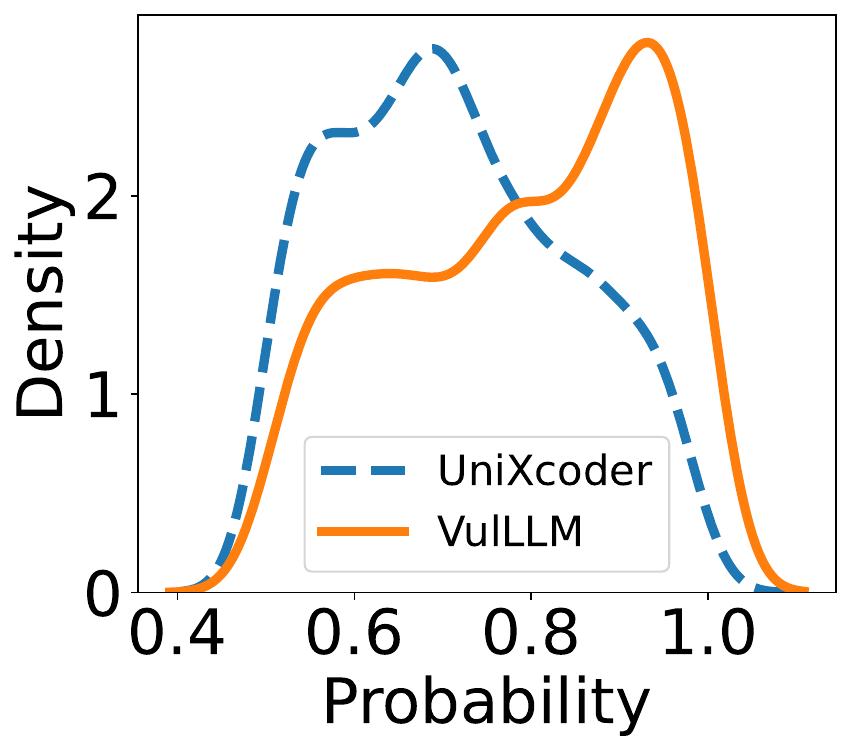}
    \caption{DiverseVul} \label{fig:density_DiverseVul}
\end{subfigure}
\begin{subfigure}{0.23\textwidth}  
  \centering
    \includegraphics[width=1\linewidth]{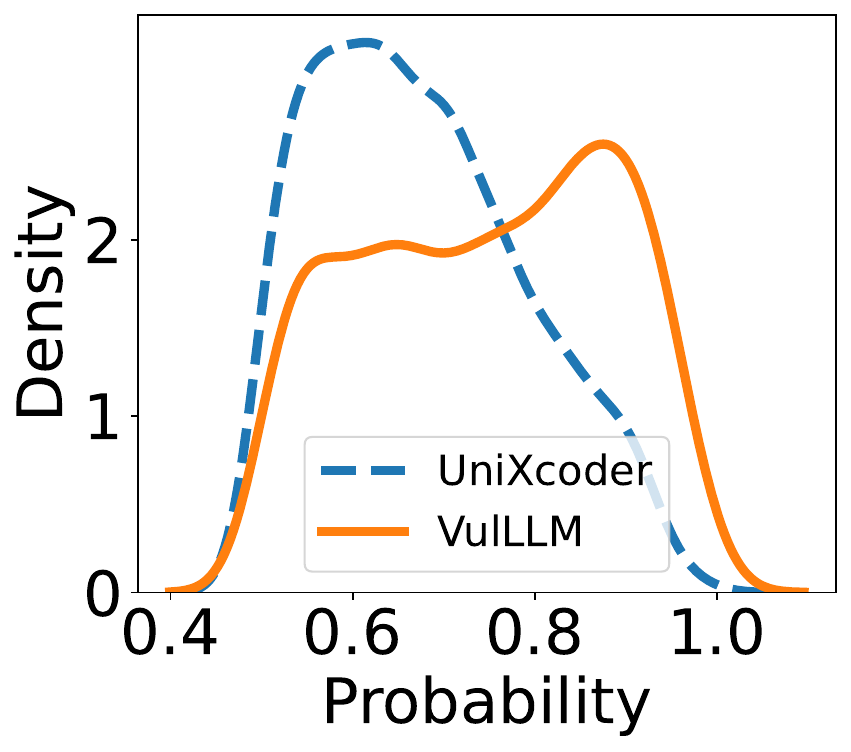}
    \caption{ReVeal} \label{fig:density_ReVeal}
\end{subfigure}
\caption{Probability density of the DiverseVul and ReVeal from VulLLM and UniXcoder}
\label{fig:density}
\end{figure}

%% file: tables/ablation.tex
\begin{table*}[t]
\centering
\small
\vspace{-2mm}
\renewcommand{\arraystretch}{1.0}
\setlength\tabcolsep{6.0pt}
\begin{tabular}{l|cc|cccc|c}
\toprule
Methods &DiverseVul &Devign &BigVul &CVEfixes &ReVeal &Juliet &Total Avg         \\ 
\midrule  
VulLLM        &70.99 &71.63 &64.42 &61.92 &64.52 &65.77 &66.54  \\
\midrule  
w/o DA        &68.89\textcolor{red}{$\downarrow$} &65.08\textcolor{red}{$\downarrow$} &59.80\textcolor{red}{$\downarrow$} &61.46\textcolor{red}{$\downarrow$} &65.65\textcolor{blue}{$\uparrow$} &66.48\textcolor{blue}{$\uparrow$} &64.56\textcolor{red}{$\downarrow$}  \\
w/o MT        &66.62\textcolor{red}{$\downarrow$} &66.56\textcolor{red}{$\downarrow$} &53.42\textcolor{red}{$\downarrow$} &59.61\textcolor{red}{$\downarrow$} &52.87\textcolor{red}{$\downarrow$} &63.00\textcolor{red}{$\downarrow$} &60.35\textcolor{red}{$\downarrow$}   \\
w/o DA\&MT     &70.08\textcolor{red}{$\downarrow$} &64.06\textcolor{red}{$\downarrow$} &56.22\textcolor{red}{$\downarrow$} &59.47\textcolor{red}{$\downarrow$} &60.31\textcolor{red}{$\downarrow$} &70.88\textcolor{blue}{$\uparrow$} &63.50\textcolor{red}{$\downarrow$}   \\
\bottomrule
\end{tabular}
\caption{\label{tab:ablation}Results of ablation study. The abbreviations ``DA'' and ``MT'' refer to the data augmentation and multi-task learning, respectively. \textcolor{red}{$\downarrow$}(\textcolor{blue}{$\uparrow$}) indicates that the performance relative to the complete VulLLM decreases (increases)}
\vspace{-3mm}
\end{table*}

%% file: tables/max_length.tex
\begin{table}[htb]
\centering
\small
\vspace{-2mm}
\renewcommand{\arraystretch}{1.1}
\setlength\tabcolsep{5.0pt}
\begin{tabular}{rr|cc|c}
\toprule
Length &Aux Data &DiverseVul &Reveal &Total Avg         \\ 
\midrule  
512          &694 &70.99 &64.52 &66.54  \\
\midrule  
1,024        &1,509 &69.66  &64.36 &66.76  \\
2,048        &2,138 &68.68  &67.00 &66.89  \\

\bottomrule
\end{tabular}
\caption{\label{tab:example}The F1 score of VulLLM under varying numbers of auxiliary task samples}
\vspace{-3mm}
\end{table}

%% file: figures/ranks.tex
\begin{figure}[htbp]
\centering
    \includegraphics[width=0.9\linewidth]{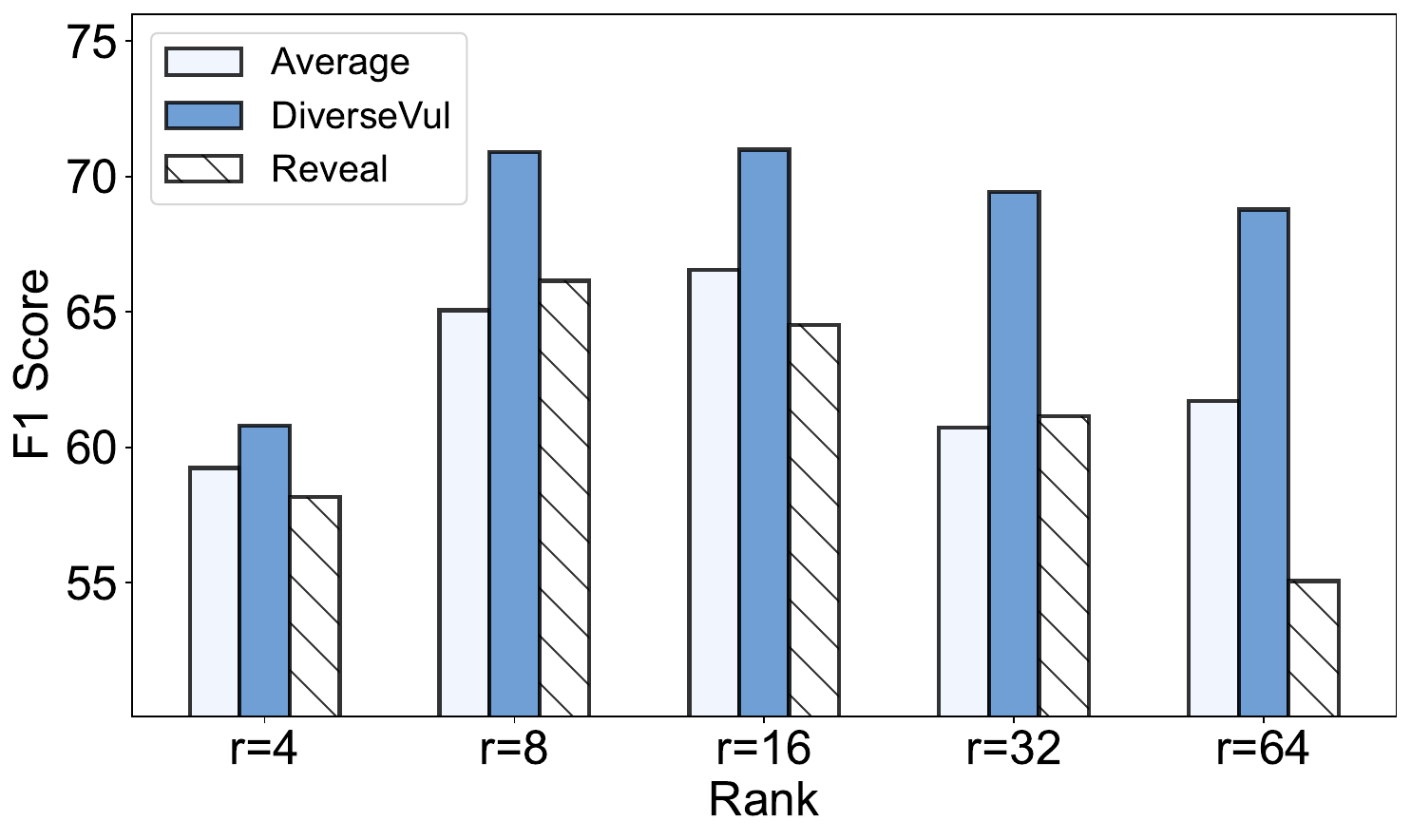}
\caption{Average F1 score for six datasets and F1 scores for DiverseVul and Reveal at different ranks}
\label{fig:ranks}
\end{figure}

%% file: conclusion.tex
\section{Conclusion}
In this paper, we introduce VulLLM, a novel framework for code vulnerability detection utilizing LLMs. 
By innovatively integrating a vulnerability interpretation task into our multi-task learning framework alongside data augmentation strategies, 
we significantly enhance the LLM's capability to detect code vulnerabilities. 
This combination not only improves detection accuracy but also enriches the model's understanding of the context and rationale behind vulnerabilities.
Extensive evaluations conducted on six diverse and comprehensive datasets demonstrate that VulLLM surpasses existing approaches in terms of effectiveness, generalization, and robustness.
Further validation through ablation study confirms the critical role of multi-task learning and data augmentation in boosting VulLLM's performance. 

%% file: limitation.tex
\section*{Limitations}
Due to resource limitations, our experiments are conducted on LLMs with size of 7B, 13B, and 15B, utilizing the parameter-efficient fine-tuning approach LoRA. This approach may affect the final performance. 
Additionally, the acquisition of vulnerability interpretations is contingent upon the capabilities inherent in the LLMs. 
To mitigate this limitation, we employ the SOTA LLM, GPT-4, in conjunction with CoT-SV for generating interpretations. 
However, there remains a potential for bias in these vulnerability explanations, particularly when dealing with code that involve complex vulnerability contexts.

%% file: acknowledgements.tex
\section*{Acknowledgements}
We sincerely thank all anonymous reviewers for their valuable comments. 
This work was supported by the Major Program (JD) of Hubei Province (No.2023BAA024), the National Natural Science Foundation of China (Grant No. 62372193), and the Young Elite Scientists Sponsorship Program by CAST (Grant No. 2021QNRC001).

%% file: appendix.tex
\section{Example of Multi-task Data}\label{sec:cot-sv}
For the vulnerable code as shown in Figure~\ref{fig:feature}, when only the prompt from \textbf{Step 1} in Section~\ref{sec:interpretation} is used, GPT-4's response is: \textit{``Buffer Overrun: The function relies on the behavior of ff\_subtitles\_next\_line(b). If this function fails to correctly handle the end of the buffer, it could result in b pointing beyond the end, causing a buffer overrun.''} However, the response following the complete CoT-SV is: \textit{``The vulnerability in svg\_probe arises from how b is incremented in line 9. This, in combination with the control flow in line 8 and the dataflow/boundary check in line 14, can lead to an infinite loop under certain conditions. This scenario highlights the importance of meticulously managing pointer arithmetic and loop conditions in low-level programming, particularly in functions processing external inputs.''} 
These results indicate that when GPT-4 directly analyzes a vulnerable code, its interpretation can be imprecise and biased, and  misidentifies the example code as \textbf{buffer overrun}. In contrast, the vulnerability interpretation provided by CoT-SV is more accurate and align more closely with the CVE descriptions as illustrated in Figure ~\ref{fig:feature}, emphasizing the \textbf{infinite loop}.
Based on the vulnerability line obtained from Figure~\ref{fig:feature} and the above-mentioned vulnerability interpretation, we present the instruction fine-tuning data examples for three tasks as shown in Table~\ref{tab:data_example}.
\input{tables/data_example}

\section{Dataset details}\label{sec:dataset}
We perform undersampling on non-vulnerability functions to ensure the numbers of vulnerable and non-vulnerable samples are balanced. 
%an equal balance in the number of samples between the two label classes.
For Devign, we utilize the standard partitions provided by the dataset to create training, validation, and test sets. 
Regarding DiverseVul, which does not provide standard partitions, we randomly split them into the training, validation, and test sets with an 8:1:1 ratio, ensuring a 1:1 ratio for both classes in each set.
Subsequently, we concatenate the training, validation, and test sets of the two datasets to obtain the corresponding training and validation sets.
The final mixed dataset used for training, validation contains 23,078 and 2,864 examples, respectively.
In our work, Big-Vul, CVEfixes, ReVeal, and Juliet are solely used as testing data. 
We also perform data cleaning, deduplication, and ensure a 1:1 ratio between positive and negative samples for these datasets.
For Big-Vul and Juliet, we randomly select 10\% of the processed data for testing. 
%\civi{Will this cause potential bias? People might challenge that the selection of the 10\% results would affect the final overall results significantly.}
%\xiaohu{If the reviewer raises this concern, we can respond as follows: Firstly, random sampling is widely recognized as a foundational method for creating unbiased datasets. This approach is commonly employed in many works related to deep learning or adversarial attacks, where a subset of data is directly sampled for experimentation. Secondly, we have employed multiple datasets (DiverseVul, Devign, BigVul, CVEfixes, ReVeal, and Juliet) for testing. The diversity of the testing data, therefore, validates the generalizability of our finding beyond the specific subsets chosen from BigVul and Juliet. Additionally, we have made all our data public. This level of transparency allows peers in the research community to replicate our work or conduct further analyses, thereby verifying the robustness of our research findings.}
As for ReVeal, due to its limited size, we utilize the entire dataset for testing.
Finally, the DiverseVul, Devign, BigVul, CVEfixes, ReVeal, and Juliet used for testing respectively contain 1,532, 1,312, 1,170, 4,216, 2,028, and 3,152 examples.

\section{Baselines}\label{sec:baselines}
\textbf{Devign}~\cite{DBLP:conf/nips/ZhouLSD019}.
Devign is a classic method that utilizes GNNs for vulnerability detection.
It extracts information from mutiple dimensions of the code, encoding it into a joint graph, and employs GGNN to learn hidden layer representations.
It uses a convolutional module to extract features from nodes for graph-level classification.
Another significant contribution of Devign is the release of a large dataset collcted and manually labeled from 4 popular C language libraries.
This dataset has been widely used in subsequent related works.
In our implementation, we use the source code released in ReVeal~\cite{DBLP:journals/tse/ChakrabortyKDR22} to conduct our experiments.

\noindent\textbf{ReVeal}~\cite{DBLP:journals/tse/ChakrabortyKDR22}.
Addressing issues such as data repetition and imbalanced data samples in existing datasets, ReVeal introduced a dataset constructed through its own collection efforts and conducted a systematic evaluation on this dataset.
Additionally, ReVeal proposed a new vulnerability detection method.
It represents code as a Code Property Graph (CPG), utilizes GGNN to obtain a graph representation, and then feeds it into a Multi-Layer Perceptron (MLP) layer for vulnerability detection.

\noindent\textbf{CodeBERT}~\cite{DBLP:conf/emnlp/FengGTDFGS0LJZ20}.
CodeBERT is a pre-trained model that is based on the RoBERTa model architecture, specifically designed for understanding and generating programming languages. 
Its training data consists of both programming languages (PL) and natural languages (NL), employing masked language modeling (MLM) and replaced token detection (RTD) as pre-training tasks. 
For fine-tuning CodePTMs on vulnerability detection, we adopte the parameter settings in CodeXGLUE~\cite{DBLP:conf/nips/LuGRHSBCDJTLZSZ21}.

\noindent\textbf{GraphCodeBERT}~\cite{DBLP:conf/iclr/GuoRLFT0ZDSFTDC21}.
GraphCodeBERT extends the BERT architecture. 
In addition to the Masked Language Modeling pre-training task on both natural language and code language inputs, GraphCodeBERT allows the incorporation of the structural information of the code (i.e., dataflow). 
Correspondingly, it introduces two additional pre-training tasks: edge prediction and node alignment.
The edge prediction and node alignment tasks are designed to encourage the model to learn semantic relationships between code structures and mapping relationships between code tokens and variable representations. 
% GraphCodeBERT was pre-trained on the CodeSearchNet dataset, and its performance was evaluated on four downstream code tasks: code search, clone detection, code translation, and code refinement.
% In this paper, we utilized the default parameter settings for fine-tuning GraphCodeBERT.

\noindent\textbf{UniXcoder}~\cite{DBLP:conf/acl/GuoLDW0022}.
UniXcoder is an unified and cross-modal pre-trained programming language model based on a N-layer Transformer architecture.
The model takes a code representation which enhanced by code comments and serialized Abstract Syntax Tree as input.
UniXcoder utilizes self-attention masks to control the model's behavior between Encoder-Only, Decoder-Only, and Encoder-Decoder. 
It concurrently employs language modeling tasks corresponding to these three behaviors for pre-training the model.
Additionally, the authors introduced two pre-training tasks to learn code semantic embeddings: multi-modal contrastive learning and cross-modal generation.
% In this paper, we adopted the Encoder-Only mode to maintain consistency with the model structure of CodeBERT and GraphCodeBERT. 
% We aligned the hyperparameter settings of UniXcoder with those of CodeBERT.

\noindent\textbf{ReGVD}~\cite{DBLP:conf/icse/NguyenNNLTP22}.
ReGVD is an effective model for code vulnerability detection. 
It treats source code as token sequences to construct graphs with node features initialized by a pre-trained language model. 
By leveraging GNNs with residual connections, ReGVD enhances learning and representation capabilities. The model combines sum and max pooling for graph embedding, which is then processed through a fully-connected and softmax layer to predict vulnerabilities.

\noindent\textbf{EPVD}~\cite{DBLP:journals/tse/ZhangLHXL23}.
EPVD works by decomposing a code snippet into several execution paths, analyzing these paths using a CodePTM and a convolutional neural network (CNN) to capture both intra- and inter-path attention, and then combining these analyses to form a comprehensive code representation. 
This representation is then used by a multilayer perceptron (MLP) classifier to identify vulnerabilities. 
This method effectively addresses issues related to irrelevant information and long code snippets in traditional vulnerability detection approaches.

\section{Metrics}
\textbf{Precision} (P) is the proportion of vulnerable code correctly predicted as vulnerability among all code predicted as vulnerability. 
\textbf{Recall} (R) is the proportion of vulnerable code correctly predicted as vulnerability among all known real vulnerable code. 
F1 denotes the harmonic mean of precision and recall and is calculated as: $F1 = 2 * (P * R) / (P + R)$. 
Given that the F1 score represents the harmonic mean of precision and recall, it effectively balances the impact of both metrics. 
Utilizing the $F_1$ score allows for a more comprehensive and equitable evaluation of the performance of vulnerability detection models. 
This ensures that the system neither generates excessive false positives nor overlooks too many genuine vulnerabilities.
Consequently, in all evaluations, the $F_1$ score is employed to evaluate the performance of different models.

\section{Implementation Details}\label{sec:mplementation}
All the experiments are conducted on an Ubuntu 20.04 server
with AMD Ryzen Threadripper 3960X 24-Core Processor CPU, 128GB
of RAM, and 2 NVIDIA A800 80G GPUs.
For fine-tuning CodePTMs, the learning rate is set to 2e-5, the max length is set to 512, the batch size is set to 32, and the epoch is set to 5. 
These parameter settings are consistent with those established on the CodeXGLUE~\cite{DBLP:conf/nips/LuGRHSBCDJTLZSZ21} benchmark.
For fine-tuning Llama-2 and CodeLlama, the learning rate is set to 1e-4, the max length also set at 512, the batch size is set to 32, and and the epoch is set to 3. 
For fine-tuning StarCoder, the learning rate is set to 2e-5, the max length also set at 512, the batch size is set to 16, and and the epoch is set to 3. 
To improve training efficiency, we load all LLMs with 8-bit quantization.
We employ LoRA~\cite{DBLP:conf/iclr/HuSWALWWC22} for instruction-tuning LLMs. The specific settings for LoRA include: the rank is 16, the alpha value is set to 32.
The target modules for Llama-2 and CodeLlama are set to `q\_proj', `v\_proj', `k\_proj', and `o\_proj', while the target module for StarCoder is set to `c\_proj', `c\_attn', `q\_attn'.
The partial parameters of different LLMs vary due to the distinct model settings provided in the official code. 
We have adhered to these settings in our experiments.

\section{Adversarial Attack}\label{sec:attack}
\textbf{MHM}~\cite{DBLP:conf/aaai/ZhangLLMLJ20}. 
MHM utilizes an iterative identifier substitution method based on the Metropolis-Hastings (M-H) sampling~\cite{metropolis1953equation}. 
This attack involves randomly choosing potential replacements for local variables and then making a strategic decision to either accept or reject these substitutions. 
MHM's effectiveness in selecting adversarial examples is enhanced by utilizing both the predicted labels and their confidence scores from the target model.

\noindent\textbf{WIR-Random}~\cite{DBLP:conf/issta/ZengTZLZZ22}.
WIR-Random employs the Word Importance Rank (WIR) method to establish the order in which identifiers are substituted. 
This attack assigns a rank to each identifier based on the change in probabilities produced by the model when the identifier is renamed to ``UNK''. 
Following this ranking, WIR-Random systematically substitutes the identifiers, choosing replacements from a random pool of candidates. 

\noindent\textbf{Dead Code Insertion}.
We employ a form of dead code construction in DIP~\cite{DBLP:conf/acl/NaC023} as follows:
\texttt{char var\_2[] = ``snippet'';}
Where \texttt{var} is an identifier randomly selected from the dataset. 
To avoid the low probability of duplicate names, it is named \texttt{var\_2}.
The code snippet is also randomly obtained from the dataset. 
An example of a dead code snippet is: 
\texttt{char xpath\_2[] = ``err = sock\_do\_ioctl(net, sock, cmd, (unsigned long)\&ktv);''}.
We ensure the syntactic correctness and semantic consistency of the original code while inserting the generated dead code into random positions within the code.

%% file: tables/data_example.tex
\begin{table*}[htb]
\centering
\small
\vspace{-2mm}
\renewcommand{\arraystretch}{1.2}
\setlength\tabcolsep{6.0pt}
\begin{tabular}{p{1.8cm}|p{13cm}}
\toprule
\multicolumn{1}{c|}{Task} & Instruction Data \\
\midrule  
\tabincell{c}{Vulnerability \\ Detection} & \tabincell{l}{Instruction: Detect whether the following code contains vulnerabilities. \\
Input: \{The vulnerable code in Figure~\ref{fig:feature}\} \\ 
Output: 1 } \\
\midrule  
\tabincell{c}{Vulnerability \\ Localization} &\tabincell{l}{Instruction: Identify any security vulnerabilities in the following code, and specify the lines where they \\ occur. \\
Input: \{The vulnerable code in Figure~\ref{fig:feature}\} \\ 
Output: 9: b+= ff\_subtitles\_next\_line (b); } \\
\midrule  
\tabincell{c}{Vulnerability \\ Interpretation} &\tabincell{l}{Instruction: Analyze the following code from the perspective of whether it contains vulnerabilities.  \\
Input: \{The vulnerable code in Figure~\ref{fig:feature}\} \\ 
Output: The vulnerability in svg\_probe arises from how b is incremented in line 9. This, in combination \\ with the control flow in line 8 and the dataflow/boundary check in line 14, can lead to an infinite loop \\ under certain conditions. This scenario highlights the importance of meticulously managing pointer \\ arithmetic and loop conditions in low-level programming, particularly in functions processing external  \\inputs. } \\
\bottomrule
\end{tabular}
\caption{\label{tab:data_example}Instruction examples of different task}
\vspace{-3mm}
\end{table*}